\documentclass{mem}
\usepackage{natbib}\usepackage{txfonts}\usepackage{balance}
\usepackage{graphicx}
\usepackage[a4paper,breaklinks,dvipdfm]{hyperref}
\idline{1}{1}

\begin{document}
\def\teff{$T\rm_{eff }$}
\def\kms{$\mathrm {km s}^{-1}$}

\title{
GRB observations with Konus-\textit{WIND} experiment
}

   \subtitle{}

\author{
D.~Frederiks, 
D.~Svinkin, 
A.~Tsvetkova, 
R.~Aptekar,	
S.~Golenetskii, 
A.~Kozlova,	
A.~Lysenko, 
\and
M.~Ulanov 
}

\institute{
Ioffe~Institute, Politekhnicheskaya~26, St.~Petersburg~194021, Russia;
\email{fred@mail.ioffe.tu}
}

\authorrunning{Frederiks }

\titlerunning{GRB observations with Konus-\textit{WIND} experiment
}

\abstract{
We give a short review of gamma-ray burst (GRB) observations with the Konus-\textit{Wind} (KW) experiment, 
which has been providing a continuous all-sky coverage in the 20 keV-15 MeV band during the period from 1994 to present. 
The recent results include a systematic study of GRBs with known redshifts and a search for ultra-long GRBs in the KW archival data. 
We also discuss the KW capabilities for multi-messenger astronomy.
	
}
\maketitle{}

\section{Introduction}

Cosmic Gamma-ray Bursts (GRBs) are the brightest sources of the high-energy 
electromagnetic radiation. The physical processes behind the huge luminosity 
of the GRB sources are of fundamental interest since they provide an opportunity 
to study physical phenomena in vicinities of stellar-mass black holes. 
The high luminosities of GRBs make them detectable out to the edge of the
visible universe thus enabling to study the nature of the first stars 
and to probe the matter properties along the whole line of sight to the sources. 

In this paper, we give a short review of the GRB studies with Konus-WIND experiment  
which has been a continuous all-sky coverage in the wide 20~keV-15~MeV band for almost 25 years, from November 1994 to present. 

\begin{figure*}[t!]
\resizebox{\hsize}{0.35\hsize}{\includegraphics[clip=true]{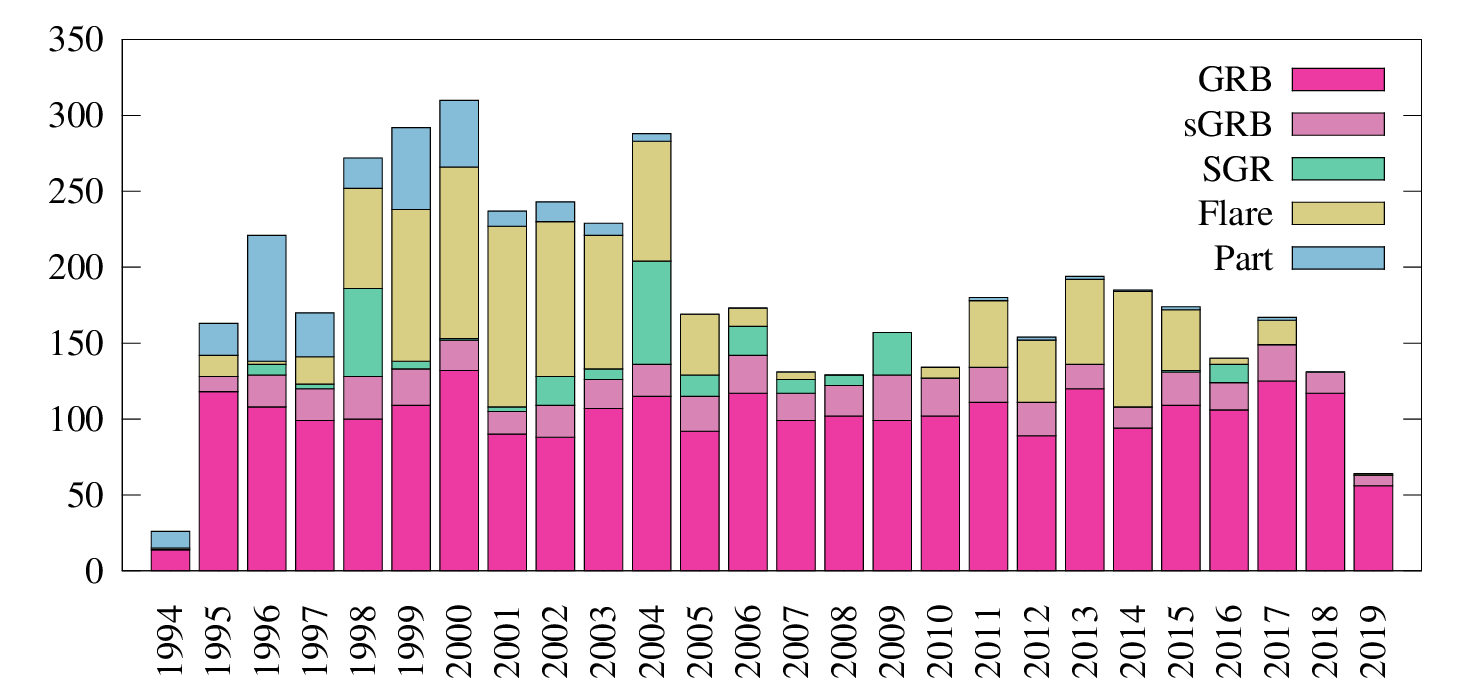}}
\caption{\footnotesize
Statistic of $\sim$4700 KW triggers from November 1994 to mid-2019.
}
\label{fig1}
\end{figure*}

\section{The KW experiment overview}
\subsection{The Instrument}
Konus-\textit{WIND} is a gamma-ray spectrometer aimed primarily at GRB and Soft Gamma Repeater (SGR) studies \citep{Aptekar95}.
It consists of two identical NaI(Tl) detectors, each with 2$\pi$ field of view, 
mounted on opposite faces of the rotationally stabilized \textit{WIND} spacecraft, 
both observing the whole sky. Each detector has an effective area of 80--160~cm$^2$ depending on
the photon energy and incident angle. The energy range of gamma-ray measurements
covers the interval from 20 keV up to 15 MeV.

The instrument has two operational modes: waiting and triggered. While in the waiting
mode, the count rates are recorded in three energy bands covering $\sim$20--1500~keV energy range with 2.944~s time resolution. 
In the triggered mode, the count rates in the three bands are recorded for $\sim$230~s with time resolution 
varying from 2 ms up to 256 ms. Simultaneously, spectral measurements are carried out in the wide 20~keV--15~MeV band. 
For a more detailed description of KW see, e.g. \cite{Svinkin16} and \cite{Tsvetkova17}.

\subsection{KW observations in 1994-2019}

Fig.~\ref*{fig1} presents yearly statistics of KW triggers in 1994--2019.
Among $>$4700 triggers to date\footnote{\url{http://www.ioffe.ru/LEA/kw/triggers/index.html}} 
$\sim$3100 are GRBs, including $\sim$500 short GRBs; 
$\sim$260 -- bright SGR bursts, including two Giant SGR flares (\citealt{Mazets99,Fred07a}); 
and $\gtrsim$1000 are Solar flares\footnote{\url{http://www.ioffe.ru/LEA/sun.html}}. 
Some notable KW GRB detections include the $\gamma$-ray coverage of the naked-eye GRB~080319B \citep{Racusin08},
the detailed study of the ultra-luminous GRB~110918A \citep{Fred13}, 
and the discovery of two extragalactic SGR candidates \citep{Fred07b,Mazets08}.

The fraction of short GRBs in the KW sample is $\sim$15\%; a detailed study of temporal and spectral 
properties of $\sim$300 short KW GRBs is given in \cite{Svinkin16}.
A recent analysis of duration and spectral-hardness distributions of $\sim$3000 KW GRBs (Svinkin et al., JPCS submitted)
suggests that about 14\% of them can be Type~I (merger-origin) and others are presumably Type~II (collapsar-origin). 

Since 2004 \textit{WIND} has been in an orbit at the L1 libration point at a distance of ∼5 lt-s. 
Far outside the Earthʼs magnetosphere, KW has the advantages over Earth-orbiting 
GRB monitors of continuous coverage, uninterrupted by Earth occultation, and a steady background. 
This makes KW the key vertex of the interplanetary network (IPN) of $\gamma$-ray detectors, 
which presently comprises six spacecraft with orbits that range from near-Earth to Martian,
and provides GRB coordinates via triangulation; the IPN localizations including KW data are regularly reported 
in Gamma-Ray Burst Coordinates Network (GCN) circulars\footnote{\url{https://gcn.gsfc.nasa.gov/gcn/}} 
and were published in several catalogs (e.g. \citealt{Palshin13,Hurley17})  
Recently, the sample of 2301 GRBs, detected by Konus-Wind in the triggered mode between 1994
and 2017 and localized by the IPN, was examined for evidence of gravitational lensing  
and no candidates for the gravitationally lensed GRBs were found at good confidence \citep{Hurley19}.

The continuous KW waiting-mode data are well-suited to the search, both blind and targeted, 
for gamma-ray transients in response to particular events, such as GRB-less supernovae (e.g. \citealt{Whitesides2017,Margutti19}), 
high-energy neutrino events, or gravitational-wave (GW) candidates \citep{Aasi14,Hurley16}. 
In the case of non-detection, KW is able of setting upper limits on soft $\gamma$-ray emission flux
from these events at several $10^{-7}$erg~cm$^{-2}$~s$^{-1}$ (10~keV--10~MeV), thus allowing to constrain an energetics of possible accompanying GRB.

\section{Recent results}
\subsection{GRBs with known redshifts}
The first part of The Konus-\textit{WIND} Catalog of Gamma-Ray Bursts with Known Redshifts
(\citealt{Tsvetkova17}, T17) presented the sample of 150 triggered GRBs (including 12 short), at $0.1 \leq z \leq 5$, 
the largest set of GRBs with known redshifts detected by a single instrument over a wide energy range.
Along with the burst durations, spectral parameters, and bolometric rest-frame energetics, 
T17 reports the updated GRB rest-frame hardness-intensity correlations,
GRB luminosity and energy-release functions and their evolutions, and the GRB formation rate (GRBFR). 

\begin{figure}[t!]
	\resizebox{\hsize}{!}{\includegraphics[clip=true]{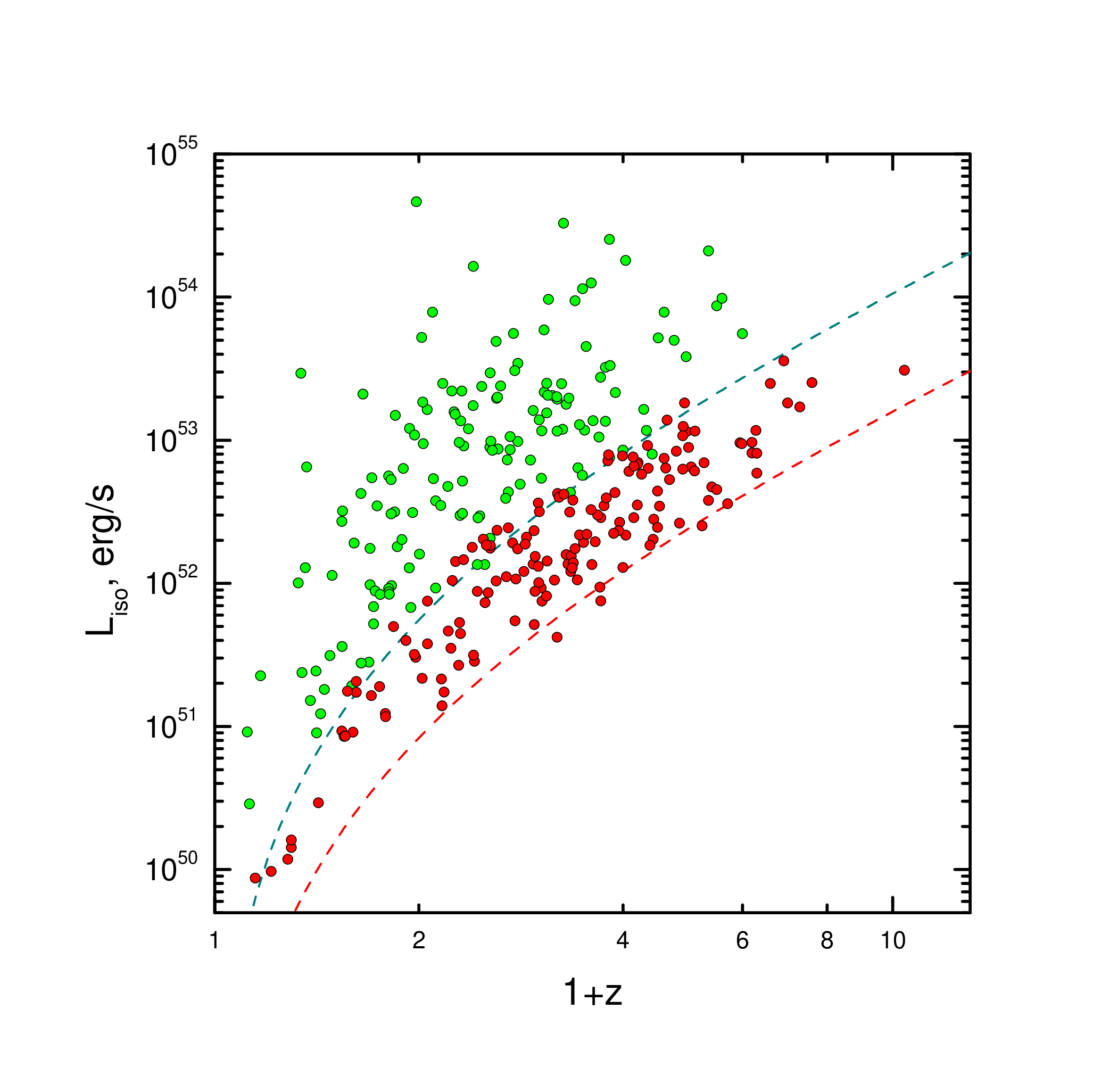}}
	\caption{
		\footnotesize
		KW GRBs with known redshifts in the $z-L_\mathrm{iso}$ plane.
		Green points show 150 triggered bursts (T17), 
		red points -- $\sim$170 waiting-mode events (preliminary).
		The dashed lines denote KW detection limits ($\sim 10^{-6}$ and $\sim 10^{-7}$erg~cm$^{-2}$~s$^{-1}$, respectively).
	}
	\label{fig2}
\end{figure}

\begin{figure}[t!]
	\resizebox{\hsize}{\hsize}{\includegraphics[clip=true]{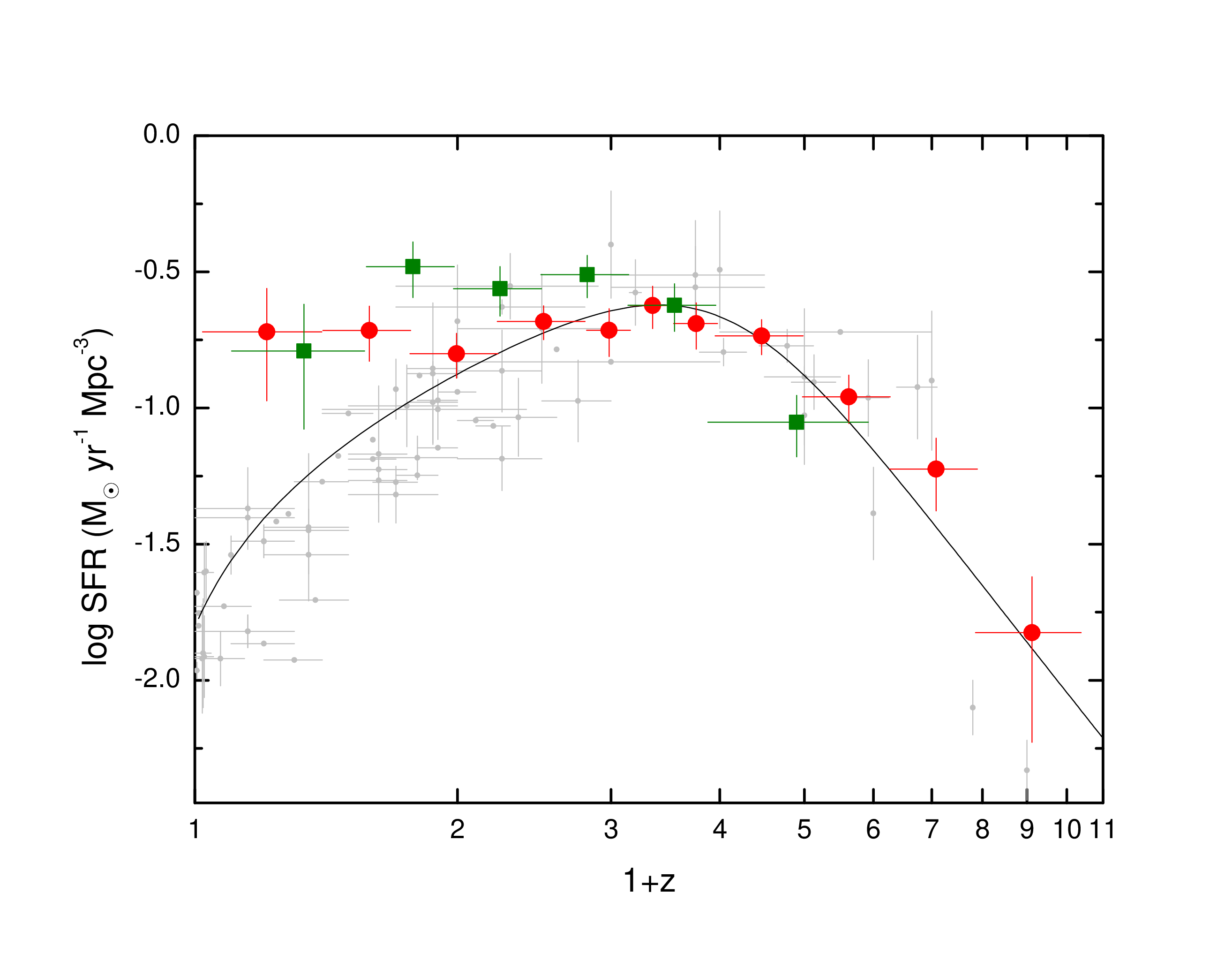}}
	\caption{
		\footnotesize
		Comparison of the GRBFR derived from the KW sample (colored points) and 
		the SFR (grey points and line, see T17 for references); 
		green points show the results from 150 triggered bursts (T17) and red points -- from the full KW sample (preliminary).
	}
	\label{fig3}
\end{figure}

The second part of the catalog, which is in preparation now, extends the T17 sample 
with $\sim$20 more triggered bursts and $\sim$170 weaker \textit{Swift} GRBs, detected by KW in the waiting mode; 
the redshift range of the extended KW sample of $>320$ GRBs is $0.04 \leq z \leq 9.4$.
Fig.~\ref*{fig2} presents the burst distribution in the $z-L_\mathrm{iso}$ plane, 
and Fig.~\ref*{fig3} shows the KW-derived GRBFR, which features a notable excess 
over star-formation rate (SFR) at $z<1$ and nearly traces the SFR at higher redshifts.   

\subsection{Search for ultra-long GRBs in the KW archival data}
\begin{figure}[t!]
	\resizebox{\hsize}{0.8\hsize}{\includegraphics[clip=true]{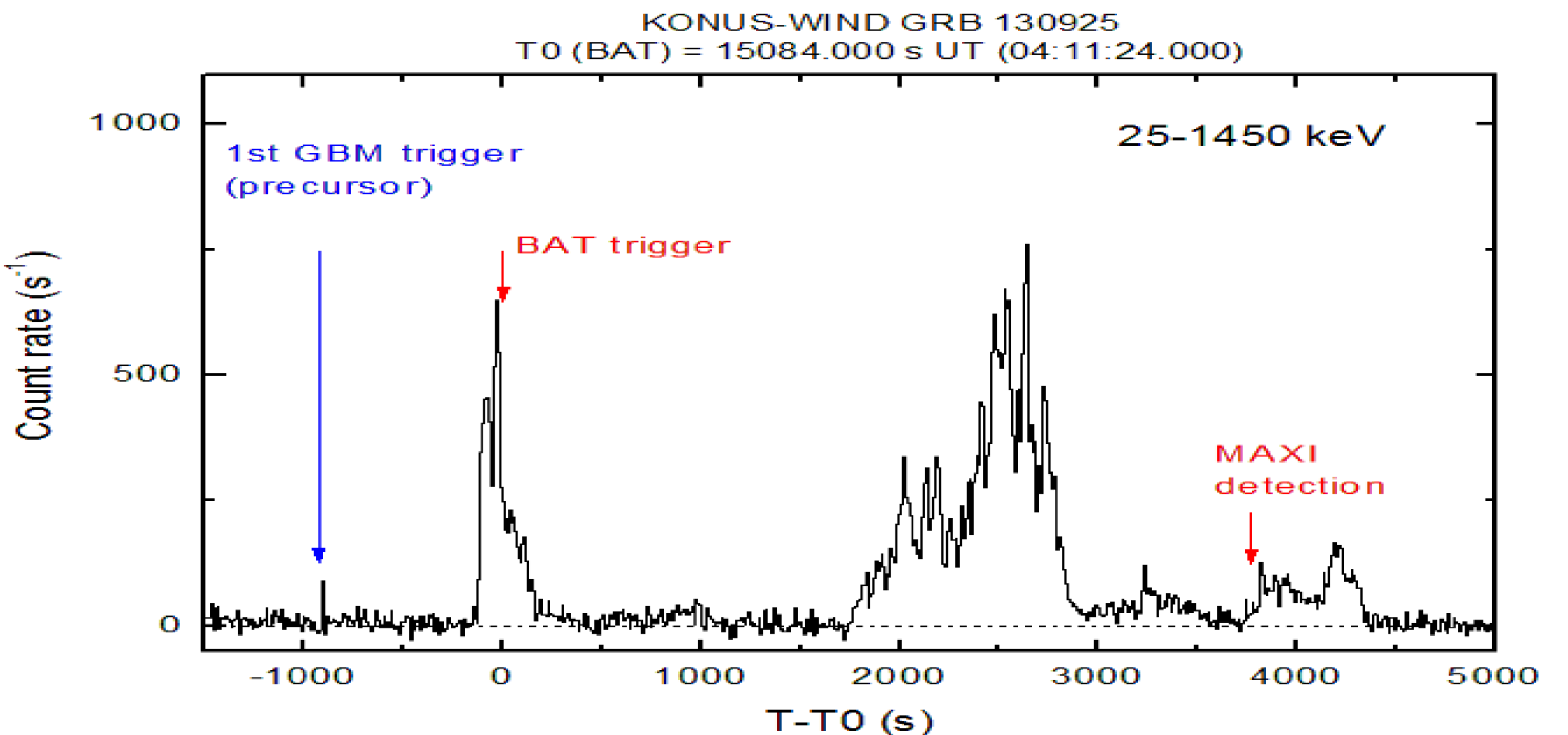}}
	\caption{\footnotesize
		Light curve of ultra-long GRB~130925A recorded by KW whole duration of the burst (black line). 
	}
	\label{fig4}
\end{figure}

Ultra-long GRBs is a special class of events with durations of kiloseconds. 
About a dozen such events are known and their exact nature is still elusive.
The continuous KW waiting-mode data provide an excellent opportunity to observe prompt emission of ultra-long
GRBs for the whole duration (Fig.~\ref*{fig4}) and to constrain their spectra and fluences in the wide energy band
(e.g. \citealt{Palshin08,Golenetskii11,Virgili13,Evans2014,Greiner2014}).

\begin{figure}[t!]
	\resizebox{\hsize}{0.8\hsize}{\includegraphics[clip=true]{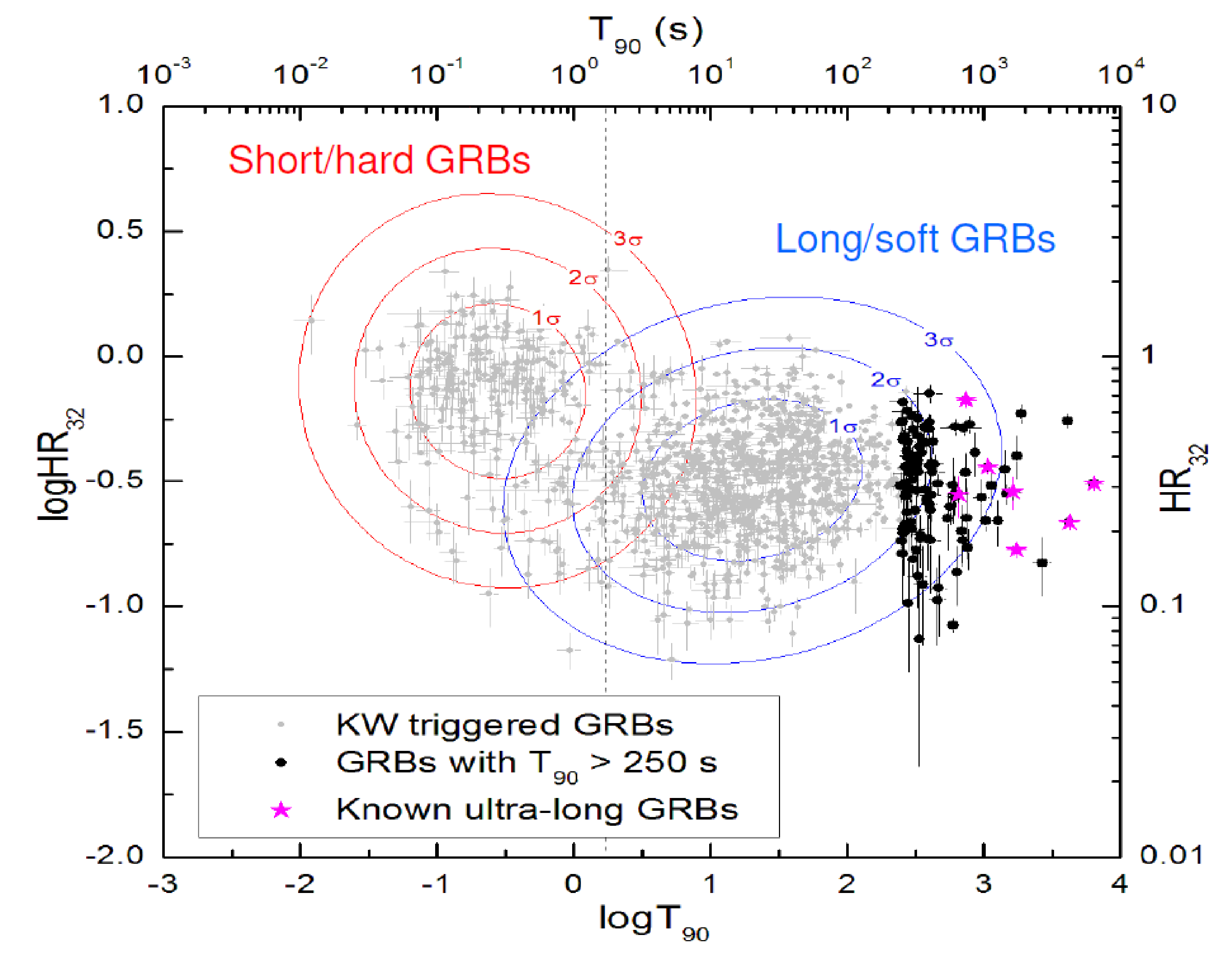}}
	\caption{
		\footnotesize
		Hardness-duration distribution of very- and ultra-long GRBs detected in the KW waiting mode (preliminary).
		The distribution of 1143 KW bright GRBs (grey points, \citealt{Svinkin16}) is shown in the background.
	}
	\label{fig5}
\end{figure}

An extensive search for hard X-ray and soft $\gamma$-ray transient events 
in the archive of KW waiting-mode observations (Kozlova et al., JPCS submitted) 
revealed $\sim$5300 confirmed GRBs and GRB candidates.
A search for "very-long" ($T_{90}>250$~s) GRBs in this sample (Svinkin et al., in preparation)
has discovered 110 such events; 13 of them, including eight previously unknown ultra-long GRBs, have full durations of $>1000$~s. 
A preliminary analysis of the sample suggests that $T_{50}$ and $T_{90}$ distributions 
of very- and ultra-long GRBs smoothly extend that of `normal' long/soft KW GRBs;
the similar behavior is implied to the two-dimensional distributions in the hardness-duration planes
(Fig.~\ref*{fig5}).  

\section{Conclusions}
We gave the short review of GRB observations with the Konus-\textit{Wind} experiment, 
which has been providing a continuous all-sky coverage in the 20 keV-15 MeV band for almost 25 years, from November 1994 to present.
The KW GRB sample is the largest to date: it comprises of $\sim$3100 triggered events, including $\sim$500 short; 
and $\sim$2200 more bursts detected in the waiting mode.
The continuous KW waiting-mode data are well-suited to the search for $\gamma$-ray transients
 in response to GRB-less supernovae, high-energy neutrino events, or GW candidates.
The recent progress obtained with the KW data includes the systematic study of 150 GRBs with known redshifts, 
the largest sample detected by a single instrument over a wide energy range; 
and the extensive search for very- and ultra-long GRBs in the KW archival data, 
which has revealed eight previously unknown ultra-long bursts and 
implied a smooth transition from 'normal' to ultra-long GRB populations. 

\begin{acknowledgements}
This work is supported by RSF grant 17-12-01378.	
\end{acknowledgements}

\bibliographystyle{aa}

\end{document}